\documentclass[10pt,aps,pra,twocolumn,showpacs,superscriptaddress,preprintnumbers,amsmath,nofootinbib]{revtex4-1}

\usepackage[utf8]{inputenc} 
\usepackage{xcolor}
\usepackage{dcolumn}
\usepackage{graphicx}

\newcommand{\m}{{|\langle m \rangle|}}
\newcommand{\mno}{{|\langle m \rangle|_\text{NO}}}
\newcommand{\mio}{{|\langle m \rangle|_\text{IO}}}
\newcommand{\hl}{{T_{1/2}^{0\nu}}}
\newcommand{\bbnn}{{(\beta\beta)_{0\nu}}}

\newcommand{\at}{\alpha_{21}}
\newcommand{\att}{\alpha_{31}}
\newcommand{\atp}{\alpha_{31}'}

\newcommand{\ms}{\Delta m^2_\odot}
\newcommand{\ma}{\Delta m^2_\text{A}}

\newcommand{\mmin}{m_\text{min}}
\newcommand{\tm}{\widetilde m}
\newcommand{\mref}{\m_0}
\newcommand{\mmaxexp}{\m^\text{max}_\text{exp}}

\newcommand{\gtap}{\ \raisebox{-.4ex}{\rlap{$\sim$}} \raisebox{.4ex}{$>$}\ }
\newcommand{\ltap}{\ \raisebox{-.4ex}{\rlap{$\sim$}} \raisebox{.4ex}{$<$}\ }

\begin{document}

\preprint{SISSA 22/2018/FISI}
\preprint{IPMU18-0103}

\title{The $\mathbf{10^{-3}}$ eV Frontier in Neutrinoless Double Beta Decay}

\author{J. T. Penedo}
\affiliation{SISSA/INFN, Via Bonomea 265, 34136 Trieste, Italy}
\author{S. T. Petcov}
\altaffiliation{Also at: Institute of Nuclear Research and Nuclear Energy,
Bulgarian Academy of Sciences, 1784 Sofia, Bulgaria.}
\affiliation{SISSA/INFN, Via Bonomea 265, 34136 Trieste, Italy}
\affiliation{Kavli IPMU (WPI), Univeristy of Tokyo, 277-8583 Kashiwa, Japan}

\date{\today}

\begin{abstract}
The observation of neutrinoless double beta decay would allow to
establish lepton number violation and the Majorana nature of neutrinos.
The rate of this process in the case of  
3-neutrino mixing is controlled by
the neutrinoless double beta decay
effective Majorana mass $\m$. For
a neutrino mass spectrum with normal ordering, 
which is favoured over the spectrum with 
inverted ordering by recent global fits,
$\m$ can be significantly suppressed.
Taking into account updated 
data on the neutrino oscillation parameters,
we investigate the conditions
under which $\m$ in the 
case of spectrum with normal ordering 
exceeds $10^{-3}~(5\times 10^{-3})$ eV: 
$\mno > 10^{-3}~(5\times 10^{-3})$ eV.
We analyse first the generic case with 
unconstrained leptonic CP violation Majorana 
phases. We show, in particular, that 
if the sum of neutrino masses is found 
to satisfy $\Sigma > 0.10$ eV, 
then $\mno > 5\times 10^{-3}$ eV
for any values of the Majorana phases.
We consider also 
cases where the values for these phases
are either CP conserving or
are in line with predictive 
schemes combining flavour 
and generalised CP symmetries.
\end{abstract}


\maketitle

%
\section{Introduction}
%
%
Despite their elusiveness, neutrinos have granted us
unique evidence for physics beyond the Standard Theory.
Observations of flavour oscillations in experiments with solar,
atmospheric, reactor, and accelerator neutrinos
(see, e.g.,~\cite{Patrignani:2016xqp}) 
imply both non-trivial mixing in the leptonic sector
and above-meV masses for at least two of the 
light neutrinos.
Neutrino oscillations, however, are blind
to the absolute scale of neutrino masses
and to the nature -- Dirac or Majorana -- 
of massive neutrinos \cite{Bilenky:1980cx,Langacker:1986jv}.

In order to uncover the possible Majorana nature of these neutral fermions,
searches for the lepton-number violating process
of neutrinoless double beta \mbox{($\bbnn$-)decay} are underway
(for recent reviews, see e.g.~\cite{Vergados:2016hso,DellOro:2016tmg}).
This decay corresponds to a transition 
between the isobars $(A,Z)$ and $(A,Z+2)$,
accompanied by the emission of two electrons but --
unlike usual double beta decay -- without the emission of
two (anti)neutrinos.
A potential observation of $\bbnn$-decay is feasible,
in principle, whenever single beta decay is energetically forbidden,
as is the case for certain even-even nuclei.
The searches for $\bbnn$-decay have a long history 
(see, e.g.,~\cite{Barabash:2011mf}).
The best lower limits on the half-lives $\hl$ of this
decay have been obtained for the isotopes of germanium-76,
tellurium-130, and xenon-136:
$\hl(^{76}\text{Ge}) > 8.0 \times 10^{25}\text{ yr}$
reported by the GERDA-II collaboration~\cite{Agostini:2018tnm},
$\hl(^{130}\text{Te}) > 1.5 \times 10^{25}\text{ yr}$
obtained from the combined results of the Cuoricino, 
CUORE-0, and CUORE experiments~\cite{Alduino:2017ehq}, and 
$\hl(^{136}\text{Xe}) > 1.07 \times 10^{26}\text{ yr}$
reached by the KamLAND-Zen collaboration~\cite{KamLAND-Zen:2016pfg},
with all limits given at the 90\% CL.

In the standard scenario where
the exchange of three Majorana neutrinos 
$\nu_i$ $(i=1,2,3)$ with masses $m_i < 10$ MeV
provides the dominant contribution to the decay rate, 
the $\bbnn$-decay rate is proportional to the 
so-called effective Majorana mass $\m$.
Given the present knowledge of neutrino oscillation data,
the effective Majorana mass is bounded from below in the case
of a neutrino mass
spectrum with inverted ordering (IO) \cite{Pascoli:2002xq},
$\mio > 1.4 \times 10^{-2}$ eV.
Instead, in the case of a spectrum with normal ordering (NO),
$\m$ can be exceptionally small:
depending on the values of the lightest neutrino mass  
and of the CP violation (CPV) Majorana phases
we can have $\mno \ll 10^{-3}$ eV
(see, e.g.,~\cite{Patrignani:2016xqp}).
Recent global analyses show a preference of
the data for NO spectrum over IO spectrum
at the $2\sigma$ CL~\cite{Capozzi:2017ipn,Esteban:2016qun}.
In the latest analysis performed in \cite{Capozzi:2018ubv} 
this preference is at $3.1\sigma$ CL.

New-generation experiments seek to probe and possibly 
cover the IO region of parameter space,
working towards the $\m \sim 10^{-2}$ eV frontier.
Aside from upgrades to the ones mentioned above,
such experiments include (see, e.g.,~\cite{Vergados:2016hso,DellOro:2016tmg}):
CANDLES ($^{48}\text{Ca}$),
{\sc Majorana} and LEGEND ($^{76}\text{Ge}$),
SuperNEMO and DCBA ($^{82}\text{Se},\,^{150}\text{Nd}$),
ZICOS ($^{96}\text{Zr}$),
AMoRE and MOON ($^{100}\text{Mo}$),
COBRA ($^{116}\text{Cd}, ^{130}\text{Te}$),
SNO+ ($^{130}\text{Te}$), and
NEXT, PandaX-III and nEXO ($^{136}\text{Xe}$).
In case these searches produce a negative result, the
next frontier in the quest for $\bbnn$-decay will 
correspond to $\m \sim 10^{-3}$ eV.

In the present article we
determine the 
conditions under which the 
effective Majorana mass in the 
case of 3-neutrino mixing 
and NO neutrino mass spectrum
exceeds the millielectronvolt value. We consider both
the generic case, where the Majorana and Dirac CPV phases
are unconstrained, as well as
a set of cases in which the CPV phases take particular values,
motivated by predictive schemes combining generalised CP and flavour symmetries.
Our study is a natural continuation and extension of the
study performed in \cite{Pascoli:2007qh}.

%
\section{The effective Majorana mass}
%
%
Taking the dominant contribution to the 
$\bbnn$-decay rate, $\Gamma_{0\nu}$,
to be due to the exchange of three Majorana neutrinos 
$\nu_i$ ($m_i < 10 \text{ MeV}$; $i=1,2,3$),
one can write the inverse of the decay half-life, 
$(T_{1/2}^{0\nu})^{-1} = \Gamma_{0\nu}\,/\ln 2$, as
\begin{equation}
\big(T_{1/2}^{0\nu}\big)^{-1} \,= \,G_{0\nu}(Q,Z) \,
\big|\mathcal{M}_{0\nu}(A,Z)\big|^2\, \m^2 \,,
\end{equation}
where $G_{0\nu}$ denotes the phase-space factor,
which depends on the $Q$-value of the nuclear transition,
and $\mathcal{M}_{0\nu}$ is the
nuclear matrix element (NME) of the decay.
The former can be computed with relatively good accuracy
whereas the latter remains the predominant
source of uncertainty
in the extraction of $\m$ from the data 
(see, e.g.,~\cite{Vergados:2016hso,Iachello:2015ejm}).

The effective Majorana mass $\m$
is given by (see, e.g.,~\cite{Bilenky:1987ty}):
\begin{equation}
\m \,=\, \left|\,\sum_{i=1}^3 \,U_{\text{e}i}^2 \,m_i\,\right|\,,
\end{equation}
with $U$ being the Pontecorvo-Maki-Nakagawa-Sakata (PMNS)
leptonic mixing matrix.
The first row of $U$ is the one
relevant for $\bbnn$-decay and reads,
in the standard parametrization~\cite{Patrignani:2016xqp},
\begin{equation}
U_{\text{e}i} = 
\begin{pmatrix}
c_{12}\, c_{13}\,, &
s_{12}\, c_{13}\, e^{i\alpha_{21}/2}\,, &
s_{13}\, e^{-i\delta}\, e^{i\alpha_{31}/2\,}
\end{pmatrix}_i 
\,.
\end{equation}
Here, $c_{ij} \equiv \cos \theta_{ij}$ and $s_{ij} \equiv \sin \theta_{ij}$,
where $\theta_{ij} \in [0,\pi/2]$ are the mixing angles,
and $\delta$ and the $\alpha_{ij}$ 
are the Dirac and Majorana CPV phases~\cite{Bilenky:1980cx},
respectively ($\delta, \alpha_{ij} \in [0,2\pi]$).

The most stringent upper limit on the effective Majorana mass 
was reported by the KamLAND-Zen collaboration. 
Using the lower limit on the half-life of $^{136}$Xe 
obtained by the collaboration and quoted in the Introduction, and
taking into account the estimated uncertainties in the NMEs 
of the relevant process, the limit reads~\cite{KamLAND-Zen:2016pfg}:
\begin{equation}
\m < (0.061-0.165)~{\rm eV}\,.
\label{eq:meffKZ}
\end{equation}

Neutrino oscillation data provides information on
mass-squared differences, but not on individual neutrino masses.
The mass-squared difference $\ms$ responsible for
solar $\nu_e$ and very-long baseline reactor 
$\bar\nu_e$ oscillations is much smaller
than the mass-squared difference $\ma$ responsible for
atmospheric and accelerator $\nu_\mu$ and $\bar\nu_\mu$
and long baseline reactor $\bar\nu_e$ oscillations,
$\ms / |\ma| \sim 1/30$.
At present the sign of $\ma$ cannot be determined
from the existing data. The two possible signs
of $\ma$ correspond to two types of neutrino mass spectrum:
$\ma > 0$ -- spectrum with normal ordering (NO), 
$\ma < 0$ -- spectrum with inverted ordering (IO).
In a widely used convention we are also going to employ, 
the first corresponds to the lightest neutrino being $\nu_1$,
while the second corresponds to the lightest neutrino being 
$\nu_3$. Combined with the fact that in this convention
$\ms \equiv \Delta m^2_{21} > 0$ we have:
\begin{itemize}
\item $m_1 < m_2 < m_3$, $\Delta m^2_{31} \equiv \ma>0$, for NO; and
\item $m_3 < m_1 < m_2$, $-\Delta m^2_{23} \equiv \ma<0$, for IO,
\end{itemize}
where $\Delta m^2_{ij} \equiv m^2_i- m^2_j$.
For either ordering, $|\ma| = \max (|m_i^2-m_j^2|)$, $i,j=1,2,3$.
We also define $\mmin \equiv m_1\, (m_3)$ in the NO (IO) case.
A NO or IO mass spectrum is additionally said to be normal 
hierarchical (NH) or inverted hierarchical (IH) if respectively
$m_1 \ll m_{2,3}$ or $m_3 \ll m_{1,2}$.
In the converse limit of relatively large $\mmin$,
$\mmin \gtap 0.1$ eV, the spectrum is said to be 
quasi-degenerate (QD) and $m_1 \simeq m_2 \simeq m_3$. 
In this last case, the distinction between
NO and IO spectra is blurred and $\ms$ and $|\ma|$
can usually be neglected with respect to $\mmin^2$.

In terms of the lightest neutrino mass, CPV phases,
neutrino mixing angles, and neutrino mass-squared differences, 
the effective Majorana mass reads:
\begin{align}
|\langle m \rangle|_\text{NO} =&\,
\bigg|\mmin\,c_{12}^2\,c_{13}^2+\sqrt{\ms + \mmin^2}\, s_{12}^2\,c_{13}^2
\,e^{i\alpha_{21}} \nonumber \\
&+\, \sqrt{\ma + \mmin^2}\, s_{13}^2\,e^{i\alpha_{31}'}\bigg|
\,,
\label{eq:mno}
\end{align}
\begin{align}
|\langle m \rangle|_\text{IO} =&\, 
\bigg|\sqrt{|\ma|-\ms+\mmin^2}\,c_{12}^2\,c_{13}^2
\nonumber \\
&+\, \sqrt{|\ma| + \mmin^2}\, s_{12}^2\,c_{13}^2 \,e^{i\alpha_{21}} \nonumber \\
&+\, \mmin\, s_{13}^2\,e^{i\alpha_{31}'}\bigg|
\,,
\label{eq:mio}
\end{align}
where we have defined $\atp \equiv \att - 2\delta$.

It proves useful to recast $\mno$ and $\mio$ given above
in the form
\begin{align}
\m =
\Big|
  \tm_1
+ \tm_2\,e^{i\alpha_{21}}
+ \tm_3\,e^{i\alpha_{31}'}
\Big|\,,
\label{eq:cast}
\end{align}
with $\tm_i > 0$ ($i=1,2,3$).
It is then clear that the effective Majorana mass is the length
of the vector sum of three vectors in the complex plane,
whose relative orientations are given by the angles $\at$ and $\atp$.

For the IO case,
taking into account the $3\sigma$ ranges of 
$\Delta m^2_{32}$, $\Delta m^2_{21}$,
$\sin^2 \theta_{12}$, and $\sin^2 \theta_{13}$
summarised in Table~\ref{tab:dataIO},
one finds that there is a hierarchy between
the lengths of the three vectors,
$\tm_3 <  0.1\, \tm_2$ and $\tm_2 < 0.6\,\tm_1$,
which holds for all values of $\mmin$.
In particular, $\tm_3 = \mmin\, s_{13}^2$ can be neglected
with respect to the other terms since $s_{13}^2 \ll \cos2\theta_{12}$.%
\footnote{It follows from the current data that 
$\cos2\theta_{12}> 0.30$ at $3\sigma$ CL.}
The above implies that extremal values of $\mio$ are obtained
when the three vectors are aligned ($\at=\atp=0$, $\mio$ is maximal)
or when $\tm_1$ is anti-aligned with $\tm_{2,3}$
($\at=\atp=\pi$, $\mio$ is minimal).
It then follows that there is a lower bound on $\mio$
for every value of $\mmin$~\cite{Pascoli:2002xq}.
This bound reads:
$\mio \gtap \sqrt{|\ma| + \mmin^2}\,c_{13}^2\cos2\theta_{12} >
\sqrt{|\ma|}\,c_{13}^2\cos2\theta_{12}$, $\mio > 1.4 \times 10^{-2}$ eV,
for variations of oscillation parameters
in their respective $3\sigma$ ranges.
In the limit of negligible $\mmin$ (IH spectrum), 
$\mmin^2 \ll {|\ma|}$,
one has $\mio \in [1.4,4.9]\times 10^{-2}$ eV.

\begin{table}[t]
\caption{\label{tab:dataIO}%
Ranges for the relevant oscillation parameters in the case of
an IO neutrino spectrum, at the $3\sigma$ CL, 
taken from the global analysis of Ref.~\cite{Capozzi:2018ubv}.
As in Table~\ref{tab:dataNO}, $\ma$ is obtained from the quantities defined in
Ref.~\cite{Capozzi:2018ubv} using the best-fit value of $\Delta m^2_{21}$.
}
\begin{ruledtabular}
\begin{tabular}{cccc}
$\dfrac{\Delta m^2_{21}}{10^{-5} \text{ eV}^2}$ & 
$\dfrac{\Delta m^2_{23}}{10^{-3} \text{ eV}^2}$ &
$\dfrac{\sin^2 \theta_{12}}{10^{-1}}$ &
$\dfrac{\sin^2 \theta_{13}}{10^{-2}}$ \\[5pt]
\colrule
\rule{0pt}{10pt}
$6.92-7.91$ & $2.38-2.58$ & $2.64-3.45$ & $1.95-2.43$
\end{tabular}
\end{ruledtabular}
\end{table}

Before proceeding to the analysis of the NO case,
let us comment on present constraints on the
absolute neutrino mass scale.
The ``conservative'' upper limit of Eq.~\eqref{eq:meffKZ}, 
$\mmaxexp = 0.165$ eV,
which is in the range of the QD spectrum,  
implies, as it is not difficult to show, the following 
upper limit on the absolute Majorana neutrino mass scale 
(i.e., on the lightest neutrino mass):
$\mmin \simeq m_{1,2,3} < 0.60$ eV,
with $\mmin \ltap \mmaxexp/(\cos2\theta_{12}-s^2_{13})$,
taking into account the $3\sigma$ ranges of $\cos2\theta_{12}$ and $\sin^2 \theta_{13}$.
Measurements of the end-point electron spectrum in tritium
beta decay experiments
constrain the combination $m_\beta \equiv \sum_i |U_{\text{e}i}|^2\,m_i$. 
The most stringent upper bounds on $m_\beta$,
$m_\beta < 2.1$ eV and $m_\beta < 2.3$ eV, both at the 95\% CL, 
are given by the Troitzk~\cite{Aseev:2011dq} and Mainz~\cite{Kraus:2004zw}
collaborations, respectively. 
The KATRIN experiment~\cite{Eitel:2005hg}
is planned to either improve this bound by an order of magnitude,
or discover $m_\beta > 0.35$ eV.
Taking into account the $3\sigma$ ranges for the relevant
mixing angles and mass-squared differences, the Troitzk bound
constrains the lightest neutrino mass to be $\mmin < 2.1$ eV.
Cosmological and astrophysical data constrain instead
the sum $\Sigma \equiv \sum_i m_i$. 
Depending on the likelihood function and data set used,
the upper limit on $\Sigma$ reported by the
Planck collaboration~\cite{Aghanim:2016yuo}
varies in the interval $\Sigma < [0.34,0.72]$ eV, 95\% CL.
Including data on baryon acoustic oscillations
lowers this bound to $\Sigma < 0.17$ eV, 95\% CL.
Taking into account the $3\sigma$ ranges for the mass-squared differences,
this last bound implies $\mmin < 0.05\,(0.04)$ eV in the NO (IO) case.
One should note that the Planck collaboration analysis is based on
the $\Lambda$CDM cosmological model. 
The quoted bounds may not apply in nonstandard cosmological scenarios
(see, e.g.,~\cite{Koksbang:2017rux}).

%
\section{The case of normal ordering}
%
%
We henceforth restrict our discussion to the effective Majorana mass $\mno$,
for which there is no lower bound.
In fact, unlike in the IO case, 
here the ordering of the lengths of the $\tm_i$ depends on the value of $\mmin$
and cancellations in $\mno$ are possible:
one risks ``falling'' inside the ``well of unobservability''.

\begin{table}[t]
\caption{\label{tab:dataNO}%
Ranges for the relevant oscillation parameters in the case of
a NO neutrino spectrum, at the $n\sigma$ ($n=1,2,3$) CLs, 
taken from the global analysis of Ref.~\cite{Capozzi:2018ubv}
(cf.~Table~\ref{tab:dataIO}).
}
\begin{ruledtabular}
\begin{tabular}{lccc}
Parameter&
$1\sigma$ range&
$2\sigma$ range&
$3\sigma$ range\\
\colrule
\rule{0pt}{10pt}%
$\Delta m^2_{21}\,/\,(10^{-5} \text{ eV}^2)$ & $7.20-7.51$ & $7.05-7.69$ & $6.92-7.91$ \\
$\Delta m^2_{31}\,/\,(10^{-3} \text{ eV}^2)$ & $2.46-2.53$ & $2.43-2.56$ & $2.39-2.59$ \\
$\sin^2 \theta_{12}\,/\,10^{-1}$ & $2.91-3.18$ & $2.78-3.32$ & $2.65-3.46$ \\
$\sin^2 \theta_{13}\,/\,10^{-2}$ & $2.07-2.23$ & $1.98-2.31$ & $1.90-2.39$
\end{tabular}
\end{ruledtabular}
\end{table}

We summarise in Table~\ref{tab:dataNO}
the $n\sigma$ ($n=1,2,3$) ranges for
the oscillation parameters relevant to $\bbnn$-decay in the NO case,
obtained in the recent global analysis of Ref.~\cite{Capozzi:2018ubv}.
Considering variations of oscillation parameters in the corresponding
$3\sigma$ ranges, for $\mmin \leq 5\times 10^{-2}$ eV there is an upper bound
$\mno \leq 5.1\times 10^{-2}$ eV (obtained for $\at=\atp=0$).
In the limit of negligible $\mmin$,
$\mmin^2 \ll {|\ma|}$,
one has $\mno \in [0.9,4.2]\times 10^{-3}$ eV.

\begin{figure}[t]
\includegraphics[width=\columnwidth]{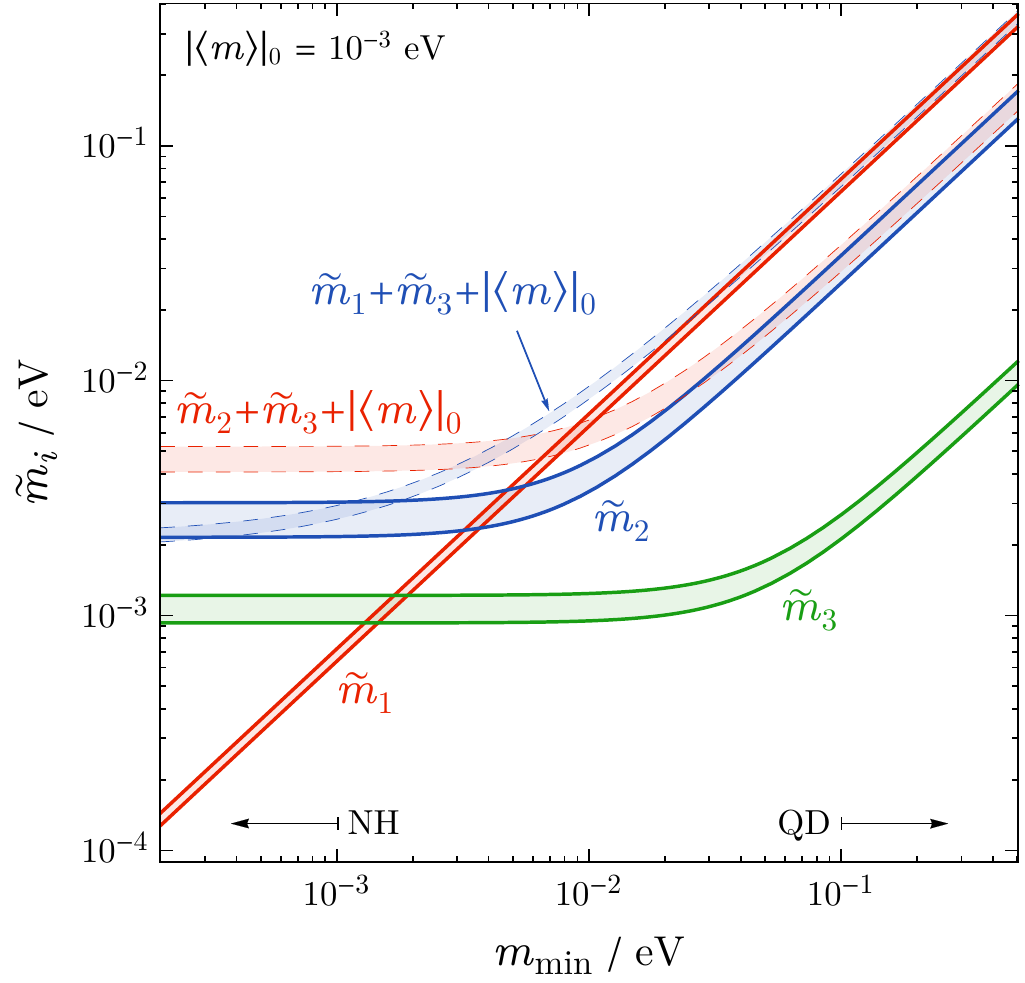}
\caption{%
Lengths $\tm_i$ of the complex vectors entering the expression of $\mno$
as a function of the lightest neutrino mass $\mmin$,
for NO spectrum.
For comparison,
the sums $\tm_1+\tm_3+\mref$ and $\tm_2+\tm_3+\mref$ are also shown (see text),
with $\mref = 10^{-3}$ eV.
Bands are obtained by varying the mixing angles and mass-squared differences
in their respective $3\sigma$ ranges (see Table~\ref{tab:dataNO}).
See text for details.
}
\label{fig:tm}
\end{figure}

From inspection of Eqs.~\eqref{eq:mno} and \eqref{eq:cast},
the vector lengths explicitly read
$\tm_1 = \mmin\,c_{12}^2\,c_{13}^2$,
$\tm_2 = \sqrt{\ms + \mmin^2}\, s_{12}^2\,c_{13}^2$, and
$\tm_3 = \sqrt{\ma + \mmin^2}\, s_{13}^2$.
In Figure~\ref{fig:tm}, these lengths are plotted as
functions of $\mmin$ for $3\sigma$ variations of oscillation parameters.

The requirement of having the effective Majorana mass above
a reference value $\mref$
is geometrically equivalent to not being able to form a quadrilateral
with sides $\tm_1$, $\tm_2$, $\tm_3$, and $\mref$.
This happens whenever one of the lengths exceeds the sum of the other three.
If however $\mref > \sum_i \tm_i$, it follows that $\m \leq \sum_i \tm_i < \mref$.
Thus, for values of $\mmin$ and oscillation parameters for which
$\tm_2 > \tm_1 + \tm_3+\mref$ or $\tm_1 > \tm_2+\tm_3+\mref$ (see Figure~\ref{fig:tm})
one is guaranteed to have $\mno > \mref$
independently of the choice of CPV phases $\at$ and $\atp$.
There are instead values of $\mmin$ for which the conditions
$\tm_2 < \tm_1 + \tm_3+\mref$ and $\tm_1 < \tm_2+\tm_3+\mref$ hold
independently of the values of oscillation parameters within a given range.
In such a case, values of $\at$ and $\atp$ such that $\mno < \mref$ are sure to exist.

\begin{figure}[t]
\includegraphics[width=\columnwidth]{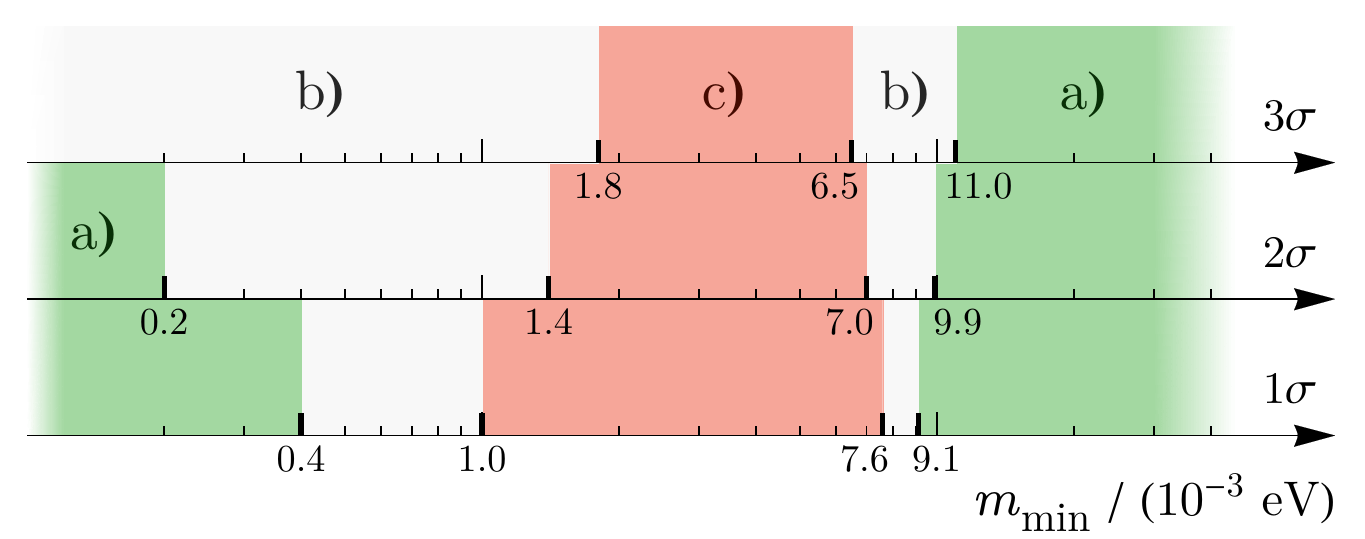}
\caption{%
Ranges of $\mmin$ for a NO spectrum and for oscillation 
parameters inside their $n\sigma$ ($n=1,2,3$) intervals 
(see Table \ref{tab:dataNO}) for which: 
in green, a) $\mno > \mref = 10^{-3}$ eV for all values of  
$\theta_{ij}$, $\Delta m^2_{ij}$, and $\alpha_{ij}^{(\prime)}$ 
from the corresponding allowed or defining intervals;
in grey, b) there exist values of $\theta_{ij}$, $\Delta m^2_{ij}$ 
from the $1\sigma$, $2\sigma$ and $3\sigma$ allowed intervals and
values of $\alpha_{ij}^{(\prime)}$ such that $\mno < \mref = 10^{-3}$ eV;
and in red, c) for all values of $\theta_{ij}$ and $\Delta m^2_{ij}$ 
from the corresponding allowed intervals
there exist values of the phases $\alpha_{21}$ and $\alpha_{31}'$
for which $\mno < \mref = 10^{-3}$ eV.
}
\label{fig:th}
\end{figure}

We summarise in Figure~\ref{fig:th} the ranges of $\mmin$
for which these different conditions apply (see caption).
We vary oscillation parameters in their respective $n\sigma$ ($n=1,2,3$) intervals
and focus on the millielectronvolt ``threshold'', $\mref = 10^{-3}$ eV.
We find that, for $3\sigma$ variations
of the $\sin^2 \theta_{ij}$ and $\Delta m^2_{ij}$,
one is guaranteed to have $\mno > 10^{-3}$ eV if 
$\mmin > 1.10 \times 10^{-2}$ eV.
This corresponds to the lower bound 
$\Sigma > 0.07$ eV on the sum of neutrino masses.
For $2\sigma$ variations, having $\mmin < 2 \times 10^{-4}$ eV
or $\mmin > 9.9 \times 10^{-3}$ eV is enough to ensure $\mno > 10^{-3}$ eV.

\begin{figure}[t]
\includegraphics[width=\columnwidth]{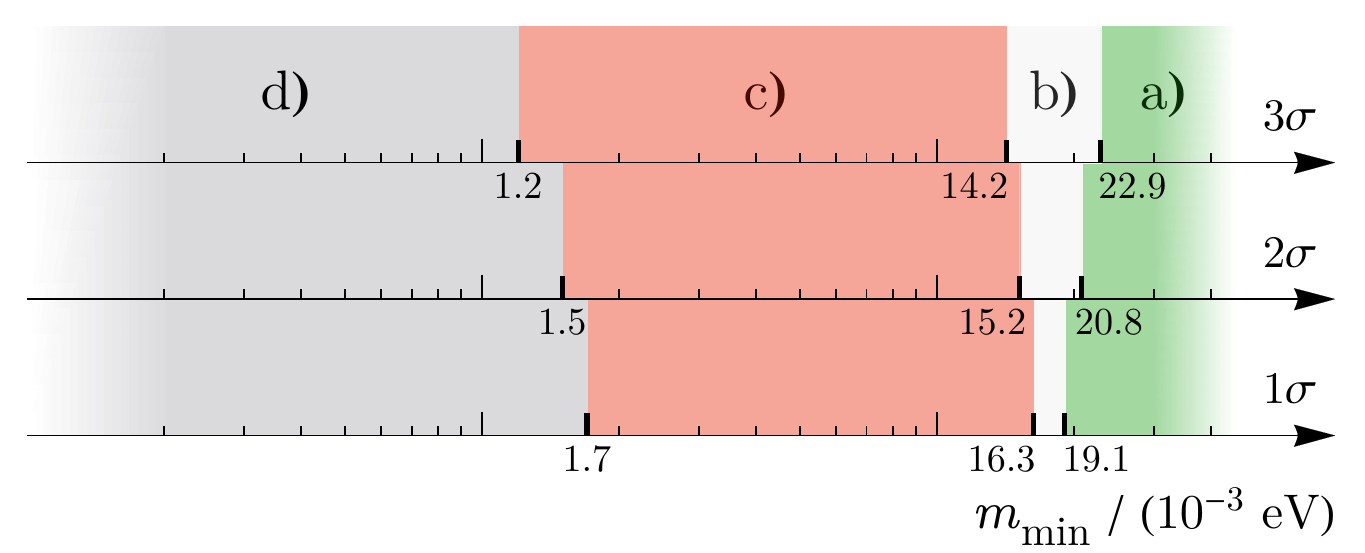}
\caption{%
The same as in Figure~\ref{fig:th}, but for the reference 
value $\mref = 5 \times 10^{-3}$ eV:
in green, a) $\mno > \mref = 5\times 10^{-3}$ eV for all values of  
$\theta_{ij}$, $\Delta m^2_{ij}$, and $\alpha_{ij}^{(\prime)}$ 
from the corresponding allowed or defining intervals;
in grey, b) there exist values of $\theta_{ij}$, $\Delta m^2_{ij}$ 
from the $1\sigma$, $2\sigma$ and $3\sigma$ allowed intervals and
values of $\alpha_{ij}^{(\prime)}$ such that $\mno < \mref = 5\times 10^{-3}$ eV;
and in red, c) for all values of $\theta_{ij}$ and $\Delta m^2_{ij}$ 
from the corresponding allowed intervals
there exist values of the phases $\alpha_{21}$ and $\alpha_{31}'$
for which $\mno < \mref = 5\times 10^{-3}$ eV.
In the darker grey ranges d) of $\mmin$, one has 
$\mno < \mref = 5\times 10^{-3}$ eV independently of 
the values of oscillation parameters and CPV phases.
}
\label{fig:th5}
\end{figure}

If one takes instead the higher value $\mref = 5\times 10^{-3}$ eV
and allows the relevant oscillation parameters 
to vary in their respective $3\sigma$ ranges,
$\mno > \mref$ is guaranteed provided $\mmin > 2.3 \times 10^{-2}$ eV,
which corresponds to the lower bound $\Sigma > 0.10$ eV on 
the sum of neutrino masses.
This lower bound on $\Sigma$ practically coincides with $\text{min}(\Sigma)$ 
in the case of IO spectrum. Thus, if $\Sigma$ is found to satisfy 
$\Sigma > 0.10$ eV, that would imply that $\m$ exceeds $5\times 10^{-3}$ eV, 
unless there exist additional contributions to the 
$\bbnn$-decay amplitude which cancel at least partially 
the contribution due to the 3 light neutrinos.
If instead $\mmin < 1.4 \times 10^{-2}$ eV, for all
($3\sigma$ allowed) values of oscillation parameters there is a choice of $\at$
and $\atp$ such that $\mno < \mref = 5 \times 10^{-3}$ eV.
These results are shown graphically in Figure~\ref{fig:th5}.

Let us briefly remark on the dependence of $\mno$ on the CPV phases. 
For the present discussion, $3\sigma$ variations of oscillation 
parameters are considered.
For all values of $\atp$ and $\epsilon>0$ there exist values of $\at$ and $\mmin$
such that $\mno < \epsilon$, i.e.~such that $\mno$ is arbitrarily small.
This is a consequence of the fact that, for any fixed oscillation parameters and $\atp$,
there is always a point $\mmin^*$ at which
$|\tm_1(\mmin^*)+\tm_3(\mmin^*)\, e^{i\atp}| = \tm_2(\mmin^*)$.
Instead, there are values of $\at$ and $\epsilon>0$ for which,
independently of $\atp$ and $\mmin$, one has $\mno > \epsilon$,
i.e.~for which $\mno$ cannot be arbitrarily small.
This conclusion may be anticipated
from the graphical results of Ref.~\cite{Xing:2016ymd},
where the structure of the $\mno$ ``well'' has been studied
as a function of $\mmin$ and $\at$.
In fact, we find that
for $\at \ltap 0.81 \pi$ or $\at \gtap 1.19 \pi$,
$\mno$ cannot be zero at tree-level
since $|\tm_1+\tm_2\, e^{i\at}| > \tm_3$, strictly.

\begin{figure}[t]
\includegraphics[width=\columnwidth]{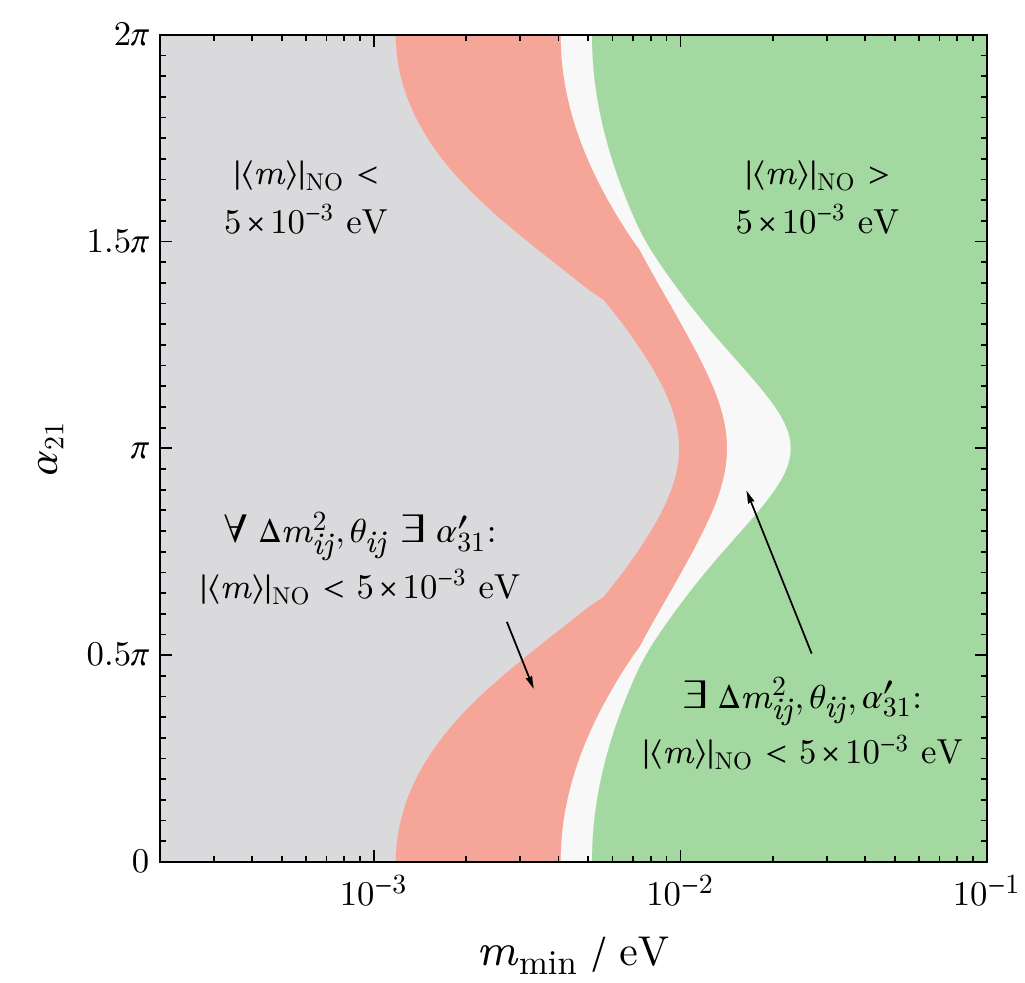}
\caption{%
Regions in the $(\mmin,\at)$ plane where different conditions on $\mno$ apply.
In the green (dark grey) region,
$\mno$ satisfies $\mno > 5 \times 10^{-3}$ eV ($\mno < 5 \times 10^{-3}$ eV)
for all values of $\theta_{ij}$, $\Delta m^2_{ij}$, and $\alpha_{13}^{\prime}$ 
from the corresponding $3\sigma$ or defining intervals.
In the red and grey regions, conditions analogous to those described in the caption of Figure~\ref{fig:th} apply and are indicated. 
This figure is to be contrasted with Figure~\ref{fig:th5},
where the dependence on $\at$ is not explicit.
}
\label{fig:plane}
\end{figure}

In Figure~\ref{fig:plane} we highlight the region of the $(\mmin,\at)$
plane in which $\mno$ is guaranteed to satisfy $\mno > 5 \times 10^{-3}$ eV,
independently of $\atp$ and of variations of oscillation 
parameters inside their $3\sigma$ ranges.

%
\section{CP and generalised CP}
%
%
Given the strong dependence of $\m$ on $\at$ and $\atp$,
some principle which determines these phases is welcome.
The requirement of CP invariance constrains the values of
the CPV phases $\at$, $\att$, and $\delta$
to integer multiples of $\pi$~\cite{Wolfenstein:1981rk,Kayser:1984ge,Bilenky:1984fg},
meaning the relevant CP-conserving values are $\at,\,\atp = 0,\,\pi$. 
Non-trivial predictions for the leptonic CPV phases may instead arise
from the breaking of a discrete symmetry combined with a generalised CP (gCP) symmetry.
We focus on schemes with large enough residual symmetry such that
the PMNS matrix depends at most on one real parameter $\theta$~\cite{Feruglio:2012cw}
and realisations thereof where the predictions for the CPV phases are unambiguous,
i.e.~independent of $\theta$.
For symmetry groups with less than 100 elements,
aside from the aforementioned CP-conserving values,
the non-trivial values $\at,\,\atp = \pi/2,\,3\pi/2$ are possible predictions~%
\cite{Ding:2013nsa,Ding:2014hva,King:2014rwa,Ding:2014ssa,Hagedorn:2014wha,%
Ding:2014ora,Ding:2015rwa}.

In what follows, we analyse the behaviour of $\mno$ and $\mio$ for each
of 16 different $(\at,\,\atp)$ pairs,
with the relevant phases taking gCP-compatible values:
$\at,\,\atp \in \{0,\,\pi/2,\,\pi,\,3\pi/2\}$.%
\footnote{Given our scope and the available literature, we find that
if $\atp = \pi/2,\, 3\pi/2$, then necessarily $\at = \pi/2,\, 3\pi/2$ is predicted.
We nevertheless take all 16 pairs of phases into consideration.}
As can be seen from Eq.~\eqref{eq:cast}, some pairs are redundant
as they lead to the same values of $\m$.
We are left with 10 inequivalent pairs of phases:
$(\at,\,\atp) =
 (\pi/2,\,0)   \sim (3\pi/2,\,0)$,
$(\pi/2,\,\pi) \sim (3\pi/2,\,\pi)$,
$(0,\,\pi/2)   \sim (0,\,3\pi/2)$,
$(\pi,\,\pi/2) \sim (\pi,\,3\pi/2)$,
$(\pi/2,\,\pi/2)\sim(3\pi/2,\,3\pi/2)$, and
$(\pi/2,\,3\pi/2)\sim(3\pi/2,\,\pi/2)$.

\begin{figure*}[t]
\includegraphics[width=1.8\columnwidth]{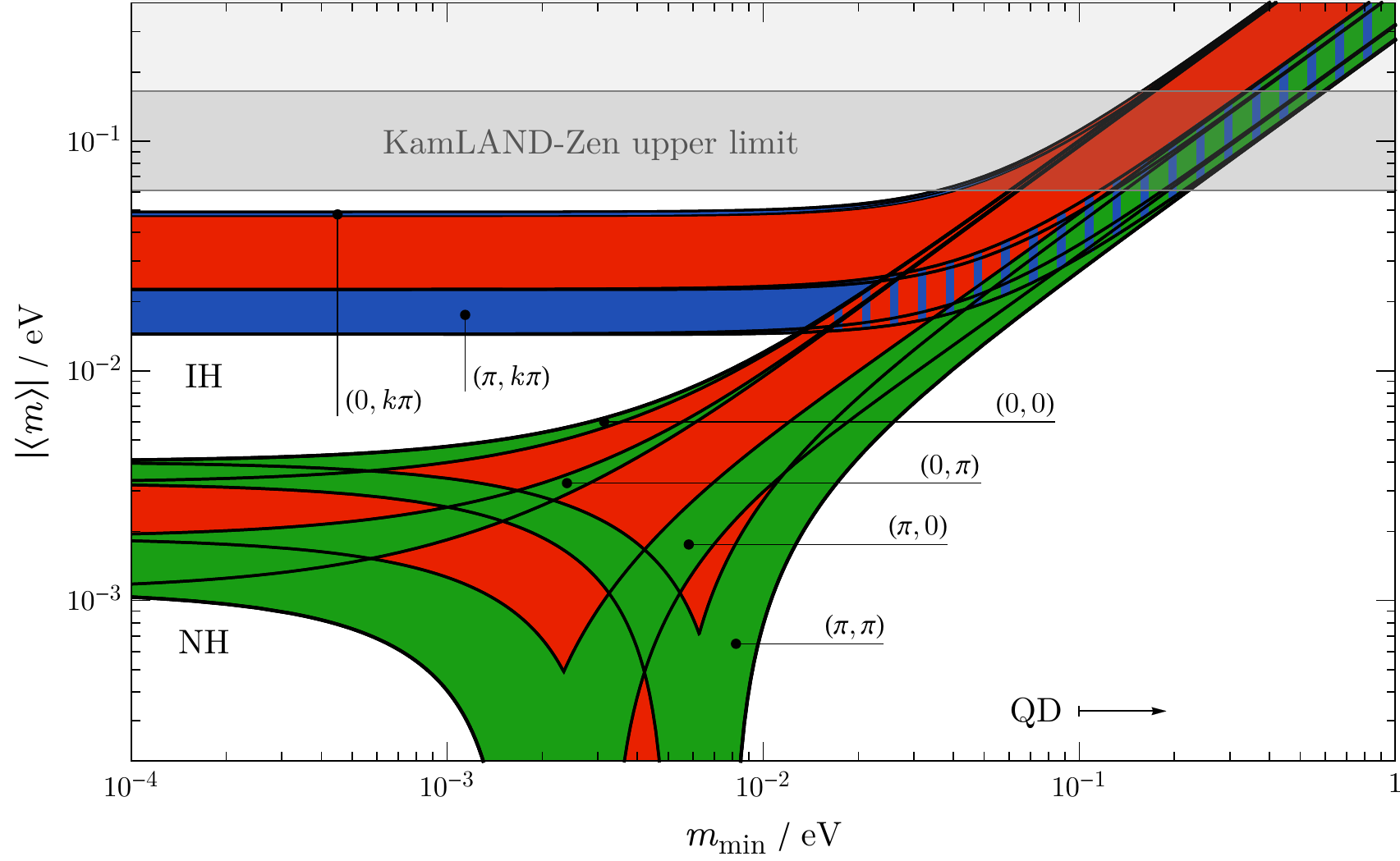}
\caption{%
The effective Majorana mass $\m$ as a function of $\mmin$, for both orderings,
allowing for variations of mixing angles
and mass-squared differences at the $2\sigma$ CL (see text).
The phases $\at$ and $\atp=\att-2\delta$ are varied in
the interval $[0,\,2\pi]$. 
Blue and green bands correspond to (the indicated, with $k=0,1$)
CP-conserving values of the phases $(\at,\,\atp)$,
for IO and NO neutrino mass spectra, respectively,
while in red regions at least one of the phases takes a CP-violating value.
Blue hatching is used to locate CP-conserving bands
in the case of IO spectrum
whenever IO and NO spectra regions overlap.
The KamLAND-Zen bound of Eq.~\eqref{eq:meffKZ} is indicated.
See also \cite{Patrignani:2016xqp,Pascoli:2007qh}.}
\label{fig:CP}
\end{figure*}

\begin{figure}[t]
\includegraphics[width=\columnwidth]{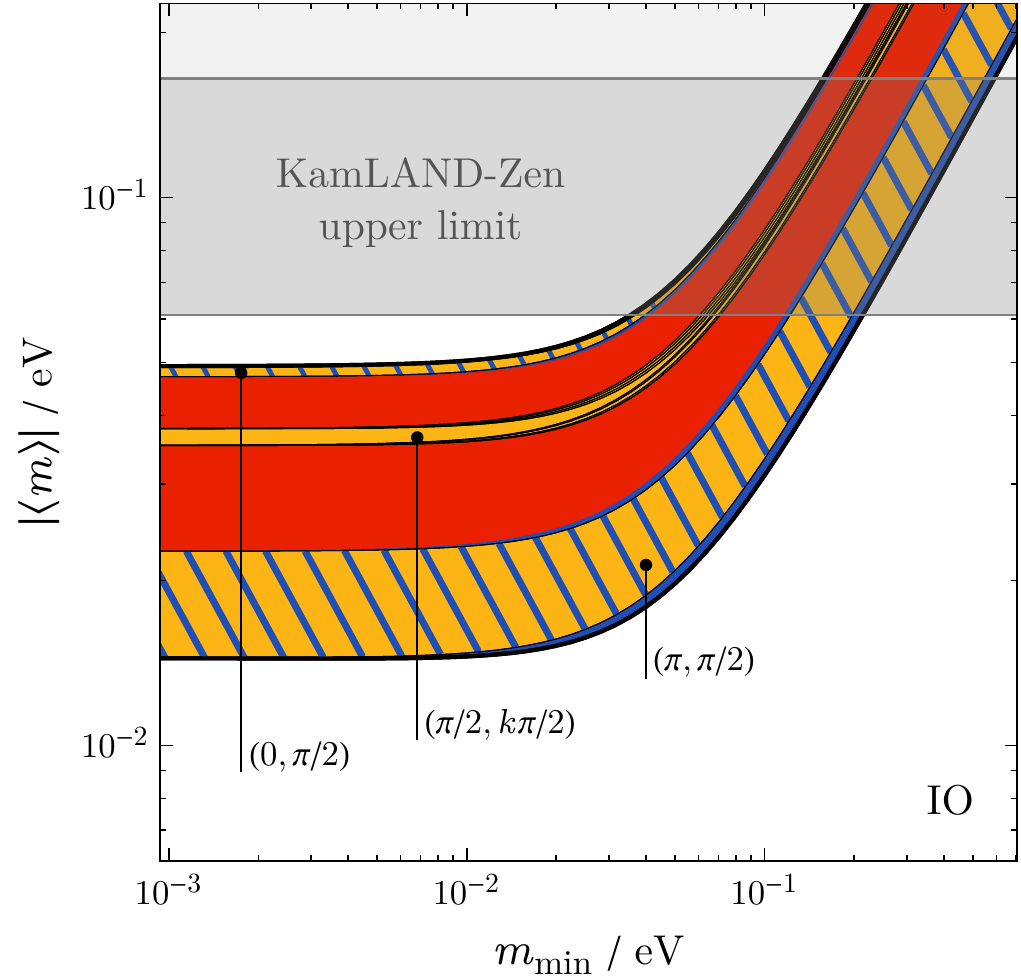}
\caption{%
The effective Majorana mass $\m$ as a function of $\mmin$,
for IO spectrum,
allowing for variations of mixing angles and mass-squared differences
at the $2\sigma$ CL (see text).
Yellow bands correspond to (the indicated, $k=0,1,2,3$) gCP-compatible
but not CP-conserving values of the phases $(\at,\,\atp)$.
Blue bands correspond to CP-conserving phases (see Figure~\ref{fig:CP})
and hatching indicates overlap with such regions,
while red regions are not gCP-compatible (for the models under consideration, see text).
The KamLAND-Zen bound of Eq.~\eqref{eq:meffKZ} is also indicated.
}
\label{fig:IO_gCP}
\end{figure}

\begin{figure*}[t]
\includegraphics[width=1.8\columnwidth]{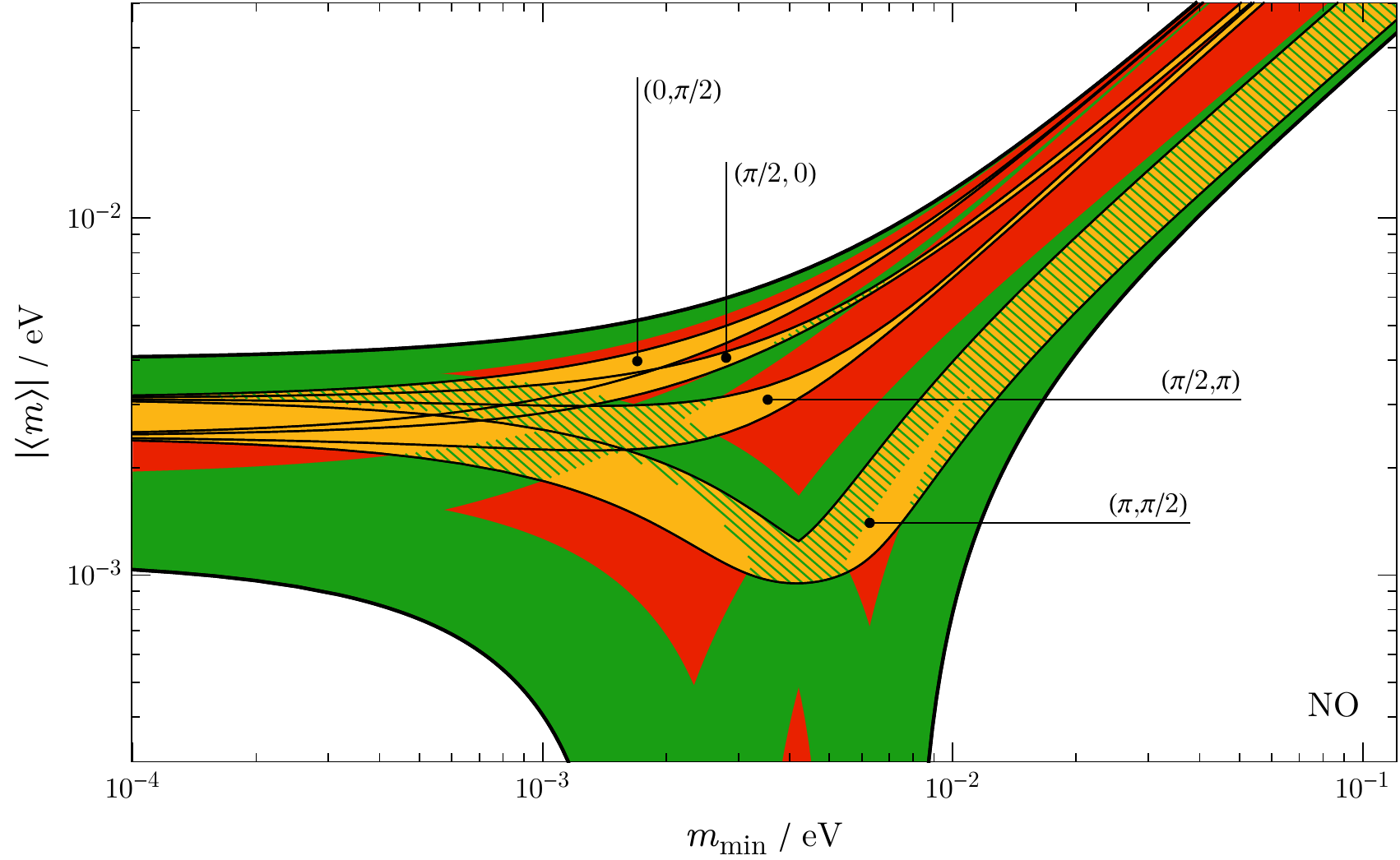}
\caption{%
The effective Majorana mass $\m$ as a function of $\mmin$,
for NO spectrum,
allowing for variations of mixing angles and mass-squared differences
at the $2\sigma$ CL (see text).
Yellow bands correspond to (the indicated) pairs $(\at,\,\atp)$ of phases,
with one CP-conserving, the other being gCP-compatible but not CP-conserving.
Green bands correspond to CP-conserving phases (see Figure~\ref{fig:CP})
and hatching indicates overlap with such regions,
while red regions are not gCP-compatible
(for the models under consideration, see text).}
\label{fig:NO_gCP1}
\end{figure*}

\begin{figure*}[t]
\includegraphics[width=1.8\columnwidth]{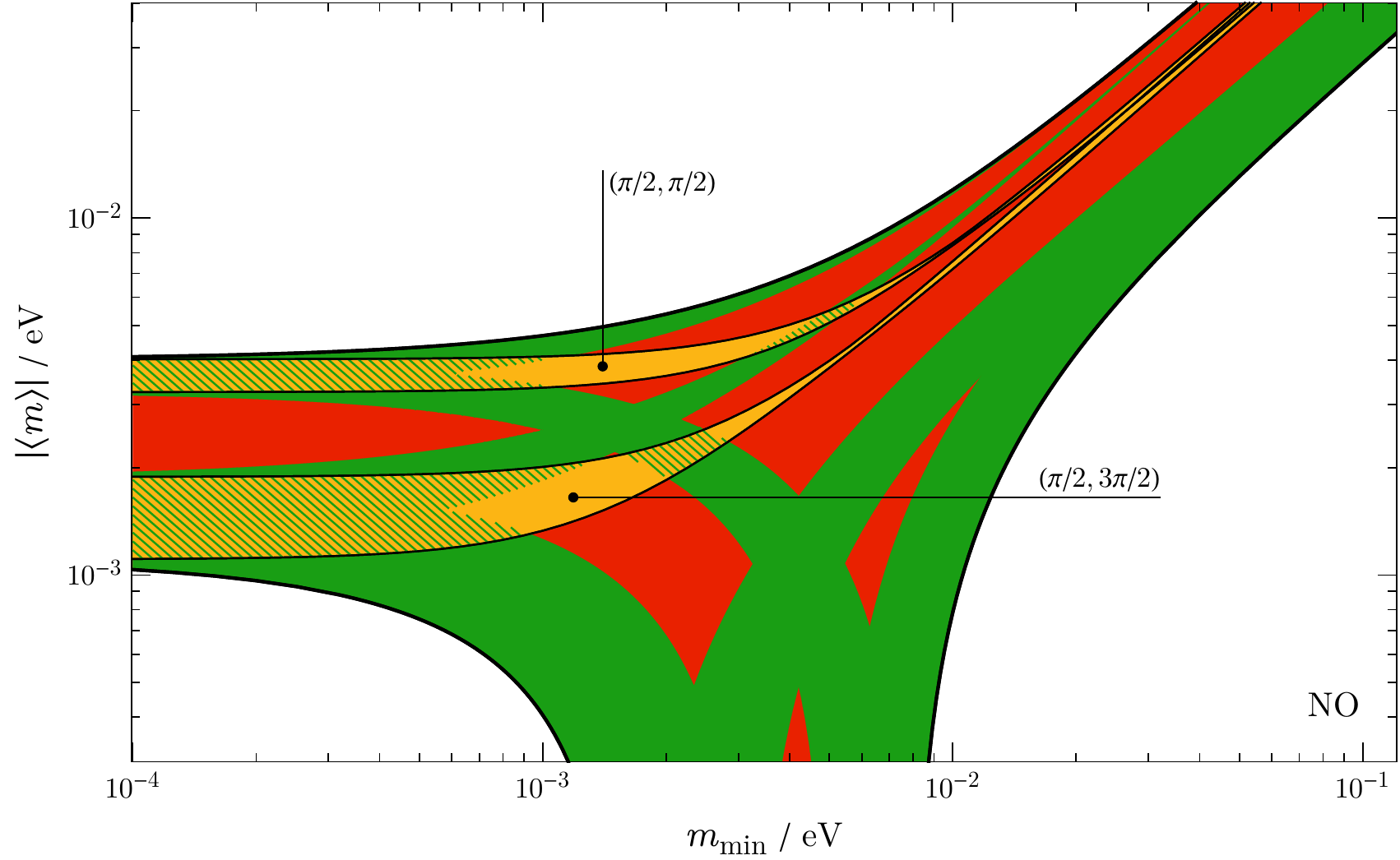}
\caption{%
The same as in Figure~\ref{fig:NO_gCP1}, with 
yellow bands corresponding to pairs $(\at,\,\atp)$
with both phases being gCP-compatible but not CP-conserving.}
\label{fig:NO_gCP2}
\end{figure*}

The $2\sigma$-allowed values of the effective Majorana mass $\m$ are presented
in Figure~\ref{fig:CP} as a function of $\mmin$, for both orderings.
Regions corresponding to different pairs $(\at,\,\atp)$ with CP-conserving phases,
$\at,\,\atp=0,\,\pi$, are singled out.
The predictions for the remaining pairs of fixed phases,
containing at least one phase which is gCP-compatible but not CP-conserving, 
are shown in Figure~\ref{fig:IO_gCP} for IO and 
in Figures~\ref{fig:NO_gCP1} and \ref{fig:NO_gCP2} for NO
(for one CP-conserving phase and for no CP-conserving phases, respectively).
Allowed values of $\m$ are found by constructing
an approximate $\chi^2$ function from the sum of the one-dimensional projections in Ref.~\cite{Capozzi:2018ubv},
and varying mixing angles and mass-squared differences
while keeping $\chi^2(\theta_{ij},\Delta m^2_{ij}) \ltap 9.72$
($2\sigma$ CL, for joint estimation of 4 parameters).

From Figures~\ref{fig:CP}\,--\,\ref{fig:NO_gCP2}, one sees that
for each value of $\mmin$ there exist values of the effective Majorana mass
which are incompatible with CP conservation.
Some of these points may nonetheless be compatible
with gCP-based predictive models.
For IO, one sees there is substantial overlap between
the bands with $(\at,\,\atp) = (0,\,0)$ and $(0,\,\pi)$,
between those of $(\pi,\,0)$ and $(\pi,\,\pi)$,
and between the four bands $(\pi/2,\,k\,\pi/2)$, with $k=0,1,2,3$.
In the case of NO, it is interesting to note that,
for a fixed, gCP-compatible but not CP-conserving pair $(\at,\,\atp)$,
$\mno$ is bounded from below at the $2\sigma$ CL,
with the lower bound at or above the meV value, $\mno \gtap 10^{-3}$ eV.
We collect in Table~\ref{tab:gCP} information
on the lower bound on $\mno$ for each pair of phases.

\begin{table}[t]
\caption{\label{tab:gCP}%
Lower bounds on $\mno$ given at the $3\sigma$ ($2\sigma$) CL,
where applicable, for different fixed values of the phases $\at$ and $\atp$. 
A tilde denotes equivalence between cases.
All bounds are given in meV.
}
\begin{ruledtabular}
\begin{tabular}{lcccc}
& \multicolumn{4}{c}{$\atp$}\\
\cline{2-5}
\rule{0pt}{10pt}%
$\at\qquad$ & $0$ & $\pi/2$ & $\pi$ & $3\pi/2$ \\
\colrule
\rule{0pt}{10pt}%
$0$      & $3.1\,(3.3)$        & $2.4\,(2.4)$    & $1.0\,(1.1)$          & $\sim(0,\,\pi/2)$ \\
$\pi/2$  & $2.4\,(2.4)$        & $3.1\,(3.3)$    & $2.1\,(2.2)$
         & $\sim(3\pi/2,\,\pi/2)$ \\
$\pi$    & no bound\footnote{$\mno > 1$ meV if $\mmin > 5.8\, (\mmin\notin [0.1,5.3])$ meV.}
         & $0.91\,(0.95)$%
\footnote{Only bounded case where $\mno$ is not strictly at or above the meV value,
for $\mmin \in [2.9,5.9]([3.2,5.3])$ meV.
$\mno > 1$ meV if e.g.~$\sin^2\theta_{13} > 2.04\times 10^{-2}$.}
         & no bound\footnote{$\mno > 1$ meV if $\mmin \notin [3.1,11.5]([3.4,10.6])$ meV.}
         & $\sim(\pi,\,\pi/2)$ \\
$3\pi/2$ & $\sim(\pi/2,\,0)$ & $1.0\,(1.1)$    & $\sim(\pi/2,\,\pi)$ & $\sim(\pi/2,\,\pi/2)$ 
\end{tabular}
\end{ruledtabular}
\end{table}

%
\section{Conclusions}
%
%
The observation of $\bbnn$-decay would allow to
establish lepton number violation and the Majorana nature of neutrinos.  
In the standard scenario of three light neutrino exchange dominance,
the rate of this process is controlled by the effective Majorana mass $\m$.
In the case of neutrino mass spectrum with inverted ordering 
(IO) the effective Majorana mass is bounded from below,
$\mio > 1.4\times 10^{-2}$ eV, where this lower bound  
is obtained using the current $3\sigma$ allowed ranges of 
the relevant neutrino oscillation parameters --
the solar and reactor neutrino mixing angles $\theta_{12}$ and $\theta_{13}$,
and the two neutrino mass-squared differences 
$\Delta m^2_{21}$ and $\Delta m^2_{23}$.
In the NO case, the effective Majorana mass $\mno$,
under certain conditions, can be exceedingly small,
$\mno \ll 10^{-2}$ eV, suppressing the $\bbnn$-decay rate.
 
Currently taking data and next-generation $\bbnn$-decay 
experiments seek to probe and possibly 
cover the IO region of parameter space,
working towards the $\m \sim 10^{-2}$ eV frontier.
In case these searches produce a negative result, the
next frontier in the quest for $\bbnn$-decay will 
correspond to $\m \sim 10^{-3}$ eV.

Taking into account updated 
global-fit data on the 3-neutrino mixing angles 
and the neutrino mass-squared differences, we have
determined the conditions under which 
the effective Majorana mass 
in the NO case $\mno$ 
exceeds the $10^{-3}$ eV ($5\times 10^{-3}$ eV) value.
The effective Majorana mass $\mno$ of interest, 
as is well known, depends 
on the solar and reactor neutrino mixing 
angles $\theta_{12}$ and $\theta_{13}$,
on the two neutrino mass-squared differences 
$\Delta m^2_{21}$ and $\Delta m^2_{31}$, on the lightest neutrino mass 
$\mmin$ as well as on the CPV Majorana phase $\at$ 
and on the Majorana-Dirac phase difference
$\atp = \att - 2\delta$.
For variations of 
$\theta_{12}$, $\theta_{13}$, 
$\Delta m^2_{21}$ and $\Delta m^2_{31}$  
in their $n\sigma$ ($n=1,2,3$) intervals,
we have determined the ranges of the 
lightest neutrino mass $\mmin$
such that (see Figures~\ref{fig:th}\,--\,\ref{fig:plane}):
\begin{enumerate}
\item[a)] $\mno > 10^{-3}~(5\times 10^{-3})$ eV independently 
of the values of $\at$ and $\atp$;
$\mno > 5\times 10^{-3}$ eV is fulfilled when $\mmin > 2.3 \times 10^{-2}$ eV
(for $3\sigma$ variations),
\item[b)] for some values of the $\theta_{ij}$ and $\Delta m^2_{ij}$
there are choices of the CPV phases $\at$ and $\atp$
such that $\mno < 10^{-3}~(5\times 10^{-3})$ eV,
\item[c)] for all values of the $\theta_{ij}$ and $\Delta m^2_{ij}$
there are choices of the CPV phases $\at$ and $\atp$ 
such that $\mno < 10^{-3}~(5\times 10^{-3})$ eV, and
\item[d)] $\mno < 5\times 10^{-3}$ eV independently 
of the values of $\at$ and $\atp$.
\end{enumerate} 
We have shown, in particular, 
that if the sum of the three neutrino masses 
is found to satisfy the lower bound $\Sigma > 0.10$ eV,
that would imply in the case of 
NO neutrino mass spectrum 
$\mno > 5\times 10^{-3}$ eV
for any values of the CPV  phases  
$\at$ and $\atp$,
unless there exist additional contributions to the 
$\bbnn$-decay amplitude which cancel at least partially 
the contribution due to the 3 light neutrinos.

We have additionally studied the predictions for $\mio$ and $\mno$
in cases where the leptonic CPV phases are fixed to particular values, 
$\at,\,\att - 2\delta \in \{0,\,\pi/2,\,\pi,\,3\pi/2\}$,
which are either CP conserving (see Figure~\ref{fig:CP})
or may arise in predictive schemes combining generalised CP
and flavour symmetries (see Figures~\ref{fig:IO_gCP}\,--\,\ref{fig:NO_gCP2},
lower bounds on the effective mass $\mno$ for such choices of phases
are given in Table~\ref{tab:gCP}).
We find that $\mno \gtap 10^{-3}$ eV for all gCP-compatible
but not CP-conserving pairs of the relevant phases.
 
The searches for lepton number non-conservation 
performed by the neutrinoless double beta decay 
experiments are part of the searches for new physics beyond 
that predicted by the Standard Theory. 
They are of fundamental importance
-- as important as the searches for 
baryon number non-conservation in the form of, 
e.g., proton decay. 
Therefore if current and next-generation $\bbnn$-decay 
experiments seeking to probe 
the IO region of parameter space
produce a negative result,
the quest for $\bbnn$-decay 
should continue towards the 
$\m \sim 5\times 10^{-3}$ eV and possibly  
the $\m \sim 10^{-3}$ eV frontier.

%
\section*{Acknowledgements}
%

We would like to thank F.~Capozzi, E.~Lisi, A.~Marrone and A.~Palazzo
for kindly sharing their data files with one-dimensional $\chi^2$ projections.
This work was supported in part by the INFN
program on Theoretical Astroparticle Physics (TASP), 
by the European Union Horizon 2020 research and innovation programme
under the  Marie Sklodowska-Curie grants 674896 and 690575 (J.T.P.~and S.T.P.),
and by the World Premier International Research Center Initiative (WPI
Initiative), MEXT, Japan (S.T.P.).

\def\bibsection{\section*{\refname}}
\bibliography{bibliography}

\end{document}